%% file: cuda_v2.tex
\documentclass[11pt]{article}
\usepackage[T1]{fontenc}
\usepackage{aas_macros}
\usepackage{mathStyle}
\usepackage[us,12hr]{datetime}
\pdfoutput=1
\usepackage{fancyhdr}
\usepackage{amsmath,amssymb,amsthm,latexsym,bbm,calc,wasysym,mathtools,empheq, yfonts,upgreek}
\usepackage[english]{babel}
\usepackage[usenames,dvipsnames]{xcolor}
\usepackage[english]{babel}
\usepackage{graphicx}
\usepackage[framemethod=tikz]{mdframed}
\usepackage[english]{babel}
\usepackage[utf8]{inputenc}
\usepackage{graphicx}
\usepackage{xcolor}
\usepackage{mathrsfs}
\usepackage{listings}
\usepackage{amsfonts}
\newcommand{\comment}[1]{}

\definecolor{oceanboatblue}{rgb}{0.0, 0.47, 0.75}
\newcommand{\CHANGE}[1]{\textcolor{black}{{#1}}}
\definecolor{rindou1}{rgb}{0.4431,0.2862,0.7960}
\definecolor{rindou2}{rgb}{0.0078,0.1215,0.4392}
\definecolor{lapis}{rgb}{0.0.0470,0.2941,0.5568}
\definecolor{mn}{rgb}{0.15, 0.35, 0.95}

\usepackage{graphicx}
\usepackage{amsfonts}
\usepackage{mathtools}
\usepackage{amssymb}
\usepackage{upgreek}
\usepackage{listings}
\usepackage{exscale,relsize}
\usepackage[makeroom]{cancel}
\usepackage{soul}
\usepackage[utf8]{inputenc}
\RequirePackage{color}
\usepackage{amsopn, amstext, wasysym}
\usepackage{colonequals}
\usepackage{rotating}
\usepackage{multirow}
\usepackage{xspace}
\usepackage{datetime}

\usepackage{tcolorbox} 
\usepackage{relsize}

\input{styles} % definition of Python code style
\newcommand{\code}[1]{\tcbox[on line,colback=pybackground,colframe=white,boxsep=3pt,left=0pt,right=0pt,top=0pt,bottom=0pt]{\lstinline[language=python,breaklines=true,breakatwhitespace=true]|#1|}}

\newcommand{\PY}{\textsc{PyTorch}\xspace}

\newcommand{\TT}{\textsc{TorchTrg}\xspace}

\newmuskip\pFqmuskip
\newcommand*\pFq[6][8]{
  \begingroup 
  \pFqmuskip=#1mu\relax
  \mathcode`\,=\string"8000
  \begingroup\lccode`\~=`\,
  \lowercase{\endgroup\let~}\pFqcomma
  {}_{#2}F_{#3}{\left[\genfrac..{0pt}{}{#4}{#5};#6\right]}
  \endgroup
}
\newcommand{\pFqcomma}{\mskip\pFqmuskip}

\usepackage[framemethod=tikz]{mdframed}

\usepackage[T1]{fontenc}
\usepackage{graphicx,scalerel}
\newmdenv[innerlinewidth=0.5pt, roundcorner=4pt,linecolor=mycolor,innerleftmargin=6pt,
innerrightmargin=6pt,innertopmargin=6pt,innerbottommargin=6pt]{mybox}

\newmdenv[innerlinewidth=0.5pt, roundcorner=4pt,linecolor=mauve,innerleftmargin=6pt,
innerrightmargin=6pt,innertopmargin=6pt,innerbottommargin=6pt]{mybox2}

\usetikzlibrary{backgrounds,fit,decorations.pathreplacing,calc}

\definecolor{navyblue}{rgb}{0.0, 0.0, 0.9}
\definecolor{rindou1}{rgb}{0.4431,0.2862,0.7960}
\definecolor{rindou2}{rgb}{0.078,0.1215,0.4392}
\definecolor{brickred}{rgb}{0.8, 0.25, 0.33}
\definecolor{BlueNCS}{rgb}{0.0, 0.53, 0.74}
\definecolor{mycolor}{rgb}{0.122, 0.435, 0.698}
\definecolor{mycolor2}{rgb}{0.02, 0.435, 0.698}
\definecolor{dkgreen}{rgb}{0,0.6,0}
\definecolor{gray}{rgb}{0.5,0.5,0.5}
\definecolor{mauve}{rgb}{0.58,0.33,0.82}
\definecolor{green2}{cmyk}{0, 1, 0.5, 0}
\definecolor{lightgreen}{cmyk}{0.2, 0, 0.2, 0.2}
\definecolor{lightgray}{cmyk}{0.1,0.2,0,0.1}
\definecolor{lightgray2}{cmyk}{0.4,0.4,0,0.8}
\definecolor{black}{cmyk}{1.0,1.0,1.0,1.0}
\definecolor{celadon}{rgb}{0.67,0.88,0.69}

\begin{document}

%%%%%%%%%%%%%%%%%%%%%%%%%%%%%%%%%%%%%%
\title{\LARGE{\textsc{GPU-Acceleration of Tensor Renormalization with PyTorch using CUDA}}} 
\author[a]{Raghav G.~Jha,}
\author[b]{Abhishek~Samlodia}
\affiliation[a]{Thomas Jefferson National Accelerator Facility, Newport News, VA 23606, USA}
\affiliation[b]{Department of Physics, Syracuse University, Syracuse NY 13244, USA}
\emailAdd{raghav.govind.jha@gmail.com}
\emailAdd{asamlodia@gmail.com}

\abstract{\\~\\ {\textsc{Abstract:} 
We show that numerical computations based on tensor renormalization group (TRG) methods can be significantly accelerated with \PY~on graphics processing units (GPUs) by leveraging NVIDIA's Compute Unified Device Architecture (CUDA). We find improvement in the runtime (for a given accuracy) and its scaling with bond dimension for two-dimensional systems. Our results establish that utilization of GPU resources is essential for future precision computations with TRG. 
}}

\maketitle
\textbf{\label{sec:0}\section{Introduction}} 
%%%%%%%%%%%%%%%%%%%%%%%%%%%%
The state-of-the-art classical method to efficiently study classical/quantum spin systems in lower dimensions is undoubtedly the tensor network method. This started with the realization that the ground state of a one-dimensional system with local Hamiltonian can be written efficiently in terms of matrix product states (MPS) which is then optimized using well-known algorithms. This idea and some of its higher dimensional generalizations are now routinely used for simulating quantum systems with low entanglement \cite{RevModPhys.93.045003}. There has been an alternate effort ~\cite{PhysRevLett.99.120601, hotrg:2012}, more natural to lattice field theory based on the Lagrangian or the partition function, known as the tensor renormalization group (TRG). This enables us to perform a version of the numerical approximation of the exact renormalization group equations to compute the Euclidean partition function by blocking the tensor network. If this blocking (coarse-graining) is applied recursively, one generates a description of the theory at increasing length scales accompanied by a corresponding flow in the effective couplings. In addition to the application of TRG to discrete spin models, where it was first introduced, it has also been used to study spin models with continuous symmetry and gauge theories in two and higher dimensions \cite{Bazavov:2019qih, Bloch:2021mjw, Kuwahara:2022ubg, Akiyama:2022eip}. We refer the interested reader to the review article~\cite{Meurice:2020pxc} to start a reference trail. 

The prospect of carrying out high-precision TRG calculations as an alternative to the standard Monte Carlo based lattice gauge computations has several motivations. The most important is the ability to study complex-action systems in the presence of finite chemical potential or topological $\theta$-term. Since the TRG algorithm does not make use of sampling techniques, they do not suffer from the sign problem~\cite{Kawauchi:2016xng, Kadoh:2019ube, Bloch:2021mjw}. However, the trade-off seems to be the fact that truncation of TRG computations (which cannot be avoided) does not always yield the correct behavior of the underlying continuum field theory. 

A major fraction of the computation time is the contraction of the tensors during successive iterations. An efficient way of doing this can lead to substantial improvements which becomes crucial when studying higher-dimensional systems. The explorations in four dimensions using ATRG \cite{atrg:2020} and HOTRG \cite{hotrg:2012} have made use of parallel CPU computing to speed up the computations and have obtained good results \cite{Akiyama:2019xzy}. 

The unreasonable effectiveness of tensors is not just restricted to de ibing the physical systems. In machine learning applications, tensors are widely used to store the higher-dimensional classical data and train the models. Due to such widespread implications of this field, several end-to-end software packages have been developed and one has now access to various scalable packages such as \textsc{TensorFlow} and \PY~which can be also be used for Physics computations. \PY~\cite{Torch2019} is a Python package that provides some high-level features such as tensor contractions with strong GPU acceleration and deep neural networks built on a reverse-mode automatic differentiation system which is an important step used in backpropagation, a crucial ingredient of machine learning algorithms. Though there have been some explorations of MPS tensor network implementations using CUDA (a parallel computing platform that allows programmers to use NVIDIA GPUs for general-purpose computing) \cite{CUDA:MPS2022_1}, it is not widely appreciated or explored in the real-space TRG community to our knowledge. CUDA provides libraries such as cuBLAS and cuDNN that can leverage tensor cores and specialized hardware units that perform fast contractions with tensors.

In this paper, we demonstrate that a simple modification of the code using \PY~with CUDA and \code{opt\_einsum} \cite{Huo2021} improves the runtime by a factor of \texttt{$\sim 12$x} with $D=89$ for the generalized XY model (described in Sec.~\ref{sec:ModelResults}). We also present results for the Ising model and the 3-state Potts model as references for the interested reader and how one can obtain state-of-the-art results in less computer time. We refer to the use of \PY~for TRG computations with CUDA as \TT and the code used to produce the results in this paper can be obtained from Ref.~\cite{Jha2023code}.

\vspace{8mm} 

%%%%%%%%%%%%%%%%%%%%%%%%%%%%
\section{\label{sec:Setup_Results}Algorithm and \TT~discussion}

We use the higher-order TRG algorithm 
based on higher-order singular value decomposition (HOSVD) of tensors. This algorithm has been thoroughly investigated in the last decade and we refer the reader to the recent review article \cite{Meurice:2020pxc} for details. The goal of this algorithm is to effectively carry out the coarse-graining of the tensor network with controlled truncation
by specifying a local bond dimension $D$ which is kept constant during the entire algorithm. \CHANGE{We show one full iteration of coarse-graining using higher-order TRG algorithm in Fig.~\ref{fig:algo1}. The first step is to combine four initial tensors (denoted $T_{0}$) as shown to construct $M_{ijkl}$. Then we combine the left and right indices to construct a matrix and take the SVD of that matrix to obtain the projector $U$. We then combine the tensors along a particular direction to obtain a coarse-grained tensor $\tilde{T}$. Then we take this tensor and perform similar steps along the orthogonal direction to obtain  $T_{1}$. This constitutes one step of coarse-graining. Doing this $N-1$ more times and then contracting the indices with periodic boundary conditions gives us the approximation to $Z$.} We show the algorithm in Fig.~\ref{fig:flow1} for the reader. The computational complexity for the higher-order TRG algorithm scales as \CHANGE{$O(D^{4d-1})$} for $d$-dimensional Euclidean systems. 
\CHANGE{The most expensive part of HOTRG computations (especially for higher dimensions) is the contraction of tensors with some fixed truncation $D$. This is needed to keep the growing size fixed to a reasonable value depending on the resources. The ratio of time complexity between tensor contraction and SVD is $O(D^{4d-7})$}. In an earlier work by one of the authors \cite{Jha:2020oik}, to perform the tensor contractions, the \code{ncon} Python library was used. There is an equivalent way of doing these contractions which has been extensively used in machine learning and is known as \code{opt\_einsum}~\cite{Smith:2018aaa}
which was used for standard CPU computations in ~\cite{Jha:2022pgy}. 
%This was used in Refs.~\cite{, Jha:2022pgy} and suggested improvements over other methods.  
In this work, we make use of additional capabilities of \code{opt\_einsum} by performing these contractions on a GPU architecture without explicitly copying any tensor to GPU device. For this purpose, a more performant backend is required which requires converting back and forth between array types. The \code{opt\_einsum} software can handle this automatically for a wide range of options such as $\textsc{TensorFlow}, \textsc{Theano}, \textsc{jax}$, and \PY. In this work, we use \PY~on NVIDIA GeForce RTX 2080 Ti. 
% nvidia-smi -q
The use of packages developed primarily for machine learning like \PY~and \textsc{TensorFlow} to problems in many-body Physics is not new. \textsc{TensorFlow} was used to study spin chains using tree tensor networks \cite{Milsted:2019ldz}  based on the software package developed in Ref.~\cite{Roberts:2019qim}. 
%We replaced the \code{ncon} in this work with \PY~backend and found that it was slower on NVIDIA GeForce RTX 2080 Ti compared to the \code{opt\_einsum} with \PY~which we have used in this work. 
However, we are not aware of any real-space tensor renormalization group algorithms which have made use of GPU acceleration with these ML/AI-based Python packages and carried out systematic study showing the improvements. Another advantage of using \PY~is the ability to carry out the automatic differentiation using: $\code{torch.tensor(T, requires\_grad = True)}$ useful in computing the derivatives similar to that in Ref.~\cite{Chen_2020}. The availability of additional GPUs also accelerates the program substantially as we will discuss in the next section. The main steps involving the conversion to the desired backend (if CUDA is available) and performing the coarse-graining step are summarized below: 
\begin{figure}
    \label{fig:algo1} 
	\centering 
	\includegraphics[width=1.0\textwidth]{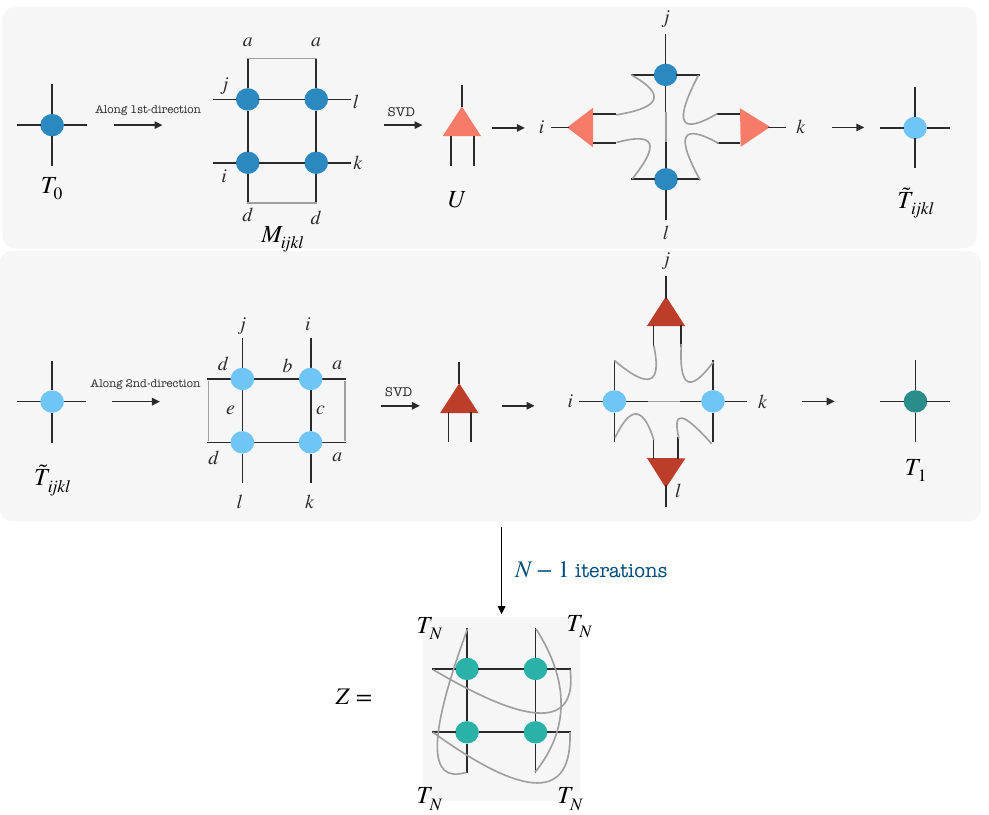}
	\caption{\label{fig:flow1}Schematic representation of the higher-order TRG implemented in this work. The diagram should be viewed from top to bottom, left to right with the first two panels denoting the coarse-graining along two directions. The first step combines four initial tensors (denoted $T_{0}$) as shown to construct $M_{ijkl}$. Then we combine the left and right indices to construct a matrix and take the SVD of that matrix to obtain the projector $U$. We then combine the tensors along a particular direction to obtain a coarse-grained tensor $\widetilde{T}$. Then with this tensor, we perform similar steps along the orthogonal direction to obtain  $T_{1}$. This constitutes one step of coarse-graining. Doing this $N-1$ times more and then contracting the indices gives us an approximation to $Z$ with periodic boundary conditions.
} 
\end{figure}

\begin{enumerate}
    \item Start with initializing all the tensors in the program as \code{torch} CPU tensors. 
    \item For tensor contractions, we use the library - \code{opt\_einsum\_torch} which utilizes GPU cores for contractions and returns a \code{torch} CPU tensor \cite{Huo2021}. 
    \item We use the linear algebra library available within \code{torch} i.e.,\code{torch.linalg} for performing SVD and other basic operations. 
\end{enumerate}
Since the tensor contractions are carried out on GPU, some fraction of the memory load on the CPU is reduced, and hence the program becomes more efficient. Furthermore, we have observed that as the architecture of the GPU improves, the computational cost improves further. We used \code{opt\_einsum} since it can significantly cut down the overall execution time of tensor network contractions by optimizing the order to the best possible time complexity and dispatching many operations to canonical BLAS or cuBLAS which provides GPU-accelerated implementation of the basic linear algebra subroutines (BLAS) \cite{Smith:2018aaa, shi2016tensor}. The order of contracting tensors is an important consideration to make in any quantum many-body computations with tensors. We revisit this issue in Appendix~\ref{sec:App} and show how they significantly differ in computation times. 
We show some code snippets with explanations below for the interested reader. The program requires three major libraries: \code{numpy,scipy,torch} which we import at the start. We also check whether we can make use of GPU i.e., whether CUDA is available. If it is available, \code{use\_cuda == True} is set for the entire computation.
 
\vspace{4mm} 
%\begin{mdframed}[backgroundcolor=celadon!6]
\begin{lstlisting}[language=Python]
import numpy as np # NumPy version 1.21.6
import scipy as sp # SciPy version 1.7.1
import torch # Torch version 1.10.1+cu102

# Import PyTorch. pip install torch usually works.  
use_cuda = torch.cuda.is_available()
# Check whether CUDA is available. If not, we do the standard CPU computation  
\end{lstlisting}
%\end{mdframed}
If CUDA is available, we print the number of devices, names, and memory and import the planner for Einstein's summation (tensor contractions). Note that 
the planner from Ref.~\cite{Huo2021} implements a memory-efficient \code{einsum} function using \PY~as backend and uses the \code{opt\_einsum} package to optimize the contraction path to achieve the minimal FLOPS. If \code{use\_cuda == False}, then we just import the basic version of the \code{opt\_einsum} package as \code{contract}. The notation for \code{contract} and CUDA based \code{ee.einsum} is similar. To compute $A_{ijkl} B_{pjql} \to C_{ipkq}$, we do: \code{C = ee.einsum('ijkl,pjql->ipkq', A, B)} with \code{use\_cuda == True} or 
\code{C = contract('ijkl,pjql->ipkq', A, B)} otherwise.

\vspace{4mm}
%\begin{mdframed}[backgroundcolor=celadon!6]
\begin{lstlisting}[language=Python]
if use_cuda:
    print('__CUDNN VERSION:', torch.backends.cudnn.version())
    print('__Number CUDA Devices:', torch.cuda.device_count())
    print('__CUDA Device Name:',torch.cuda.get_device_name(0))
    print('__CUDA Device TotalMemory[GB]:',torch.cuda.get_device_properties(0).total_memory/1e9)

    # __CUDNN VERSION: 7605
    # __Number CUDA Devices: 1
    # __CUDA Device Name: NVIDIA GeForce RTX 2080 Ti
    # __CUDA Device Total Memory [GB]: 11.554717696

    from opt_einsum_torch import EinsumPlanner
    # To install use: pip install opt-einsum-torch
    ee = EinsumPlanner(torch.device('cuda:0'), cuda_mem_limit = 0.8)

else:
    from opt_einsum import contract
    # To install use: pip install opt-einsum
    
\end{lstlisting}
%\end{mdframed}
% nvidia-smi --list-gpus
\vspace{4mm}
One thing to note is that we have to specify the CUDA memory limit for the planner. This parameter can be tuned (if needed) but we have found that a value between 0.7 and 0.85 usually works well. Note that this can sometimes limit the maximum $D$ one can employ in TRG computations. So, it should be selected appropriately if CUDA runs out of memory. The choice of this parameter and the available memory can result in errors. A representative example is: 
\begin{verbatim}
RuntimeError: CUDA out of memory. Tried to allocate 2.25 GiB (GPU 0; 10.76 GiB
total capacity; 5.17 GiB already allocated; 2.20 GiB free; 7.40 GiB reserved in 
total by PyTorch) If reserved memory is >> allocated memory try setting 
max_split_size_mb to avoid fragmentation. 
\end{verbatim}. 
We show code snippet to address this error below. 
\vspace{3mm}

\begin{lstlisting}[language=Python, upquote=true]
# Sometimes to tackle the error above, doing the below works. 
os.environ["PYTORCH_CUDA_ALLOC_CONF"] = "max_split_size_mb:<size here>"

# Tuning the cuda_mem_limit also helps. 
ee = EinsumPlanner(torch.device('cuda:0'), cuda_mem_limit = 0.7)
\end{lstlisting}
%\end{mdframed}
\vspace{4mm}
\CHANGE{The user can also completely skip the memory allocation and run without specifying. It should not affect the performance significantly.}
In implementing \TT, we explored four models that can be selected at run time by the user. The choices are:
\vspace{3mm}
%\begin{mdframed}[backgroundcolor=celadon!6]
\begin{lstlisting}[language=Python]
models_allowed = ['Ising', 'Potts','XY', 'GXY']
\end{lstlisting}
%\end{mdframed}
There are four command-line arguments: temperature $1/\beta$, bond dimension $D$, number of iterations, and the model. An example of execution is:~~\code{python 2dTRG.py 2.27 64 20 Ising}. 
Based on the model and the parameters, it constructs the initial tensor for the coarse-graining iterations to commence. It is straightforward to add other models or observables and make use of the basic CUDA setup presented here. In \TT, we have simplified the code for a non-expert to the extent that a single coarse-graining step which takes in a tensor and outputs transformed tensor and normalization factor is \code{23} lines long and can accommodate different architectures. We wrap all commands which can potentially make use of CUDA acceleration i.e., contractions etc. inside $\code{use\_cuda}$ conditional statement. We show this part of the code below:
\vspace{2mm}
%\begin{mdframed}[backgroundcolor=celadon!6]
\begin{lstlisting}[language=Python]
def SVD(t, left_indices, right_indices, D):
    T = torch.permute(t, tuple(left_indices + right_indices)) if use_cuda else np.transpose(t, left_indices + right_indices)
    left_index_sizes = [T.shape[i] for i in range(len(left_indices))]
    right_index_sizes = [T.shape[i] for i in range(len(left_indices), len(left_indices) + len(right_indices))]
    xsize, ysize = np.prod(left_index_sizes), np.prod(right_index_sizes)
    T = torch.reshape(T, (xsize, ysize)) if use_cuda else np.reshape(T, (xsize, ysize))
    U, _, _ = torch.linalg.svd(T, full_matrices=False) if use_cuda else sp.linalg.svd(T, full_matrices=False)
    size = np.shape(U)[1]
    D = min(size, D)
    U = U[:, :D]
    U = torch.reshape(U, tuple(left_index_sizes + [D])) if use_cuda else np.reshape(U, left_index_sizes + [D]) 
    return U 
    
def coarse_graining(t):
    Tfour = ee.einsum('jabe,iecd,labf,kfcd->ijkl', t, t, t, t) if use_cuda else contract('jabe,iecd,labf,kfcd->ijkl', t, t, t, t)
    U = SVD(Tfour,[0,1],[2,3],D_cut) 
    Tx = ee.einsum('abi,bjdc,acel,edk->ijkl', U, t, t, U) if use_cuda else contract('abi,bjdc,acel,edk->ijkl', U, t, t, U)
    Tfour = ee.einsum('aibc,bjde,akfc,flde->ijkl',Tx,Tx,Tx,Tx) if use_cuda else contract('aibc,bjde,akfc,flde->ijkl',Tx,Tx,Tx,Tx)
    U = SVD(Tfour,[0,1],[2,3],D_cut) 
    Txy = ee.einsum('abj,iacd,cbke,del->ijkl', U, Tx, Tx, U) if use_cuda else contract('abj,iacd,cbke,del->ijkl', U, Tx, Tx, U)
    norm = torch.max(Txy) if use_cuda else np.max(Txy)
    Txy /= norm
    return Txy, norm
\end{lstlisting}
%\end{mdframed}
\vspace{2mm}

% Sec.~\ref{sec:ModelResults}.  
%%%%%%%%%%%%%%%%%%%%%%%%%%%%

\vspace{10mm}

%%%%%%%%%%%%%%%%%%%%%%%%%%%%
\section{\label{sec:ModelResults}Models and Results} 

In this section, we show the results obtained using \TT. We first show the run time comparison on CPU and CUDA architectures for the generalized XY model, which is a deformation of the standard XY model. Then, we discuss the TRG method as applied to the classical Ising model and discuss how we converge to a desired accuracy faster. In the last part of this section, we discuss the $q$-state Potts model with $q=3$ and accurately determine the transition temperature corresponding to the continuous phase transition.  

%%%%%%%%%%%%%%%%%%%%%%%%%%%%
\subsection{\label{subsec:GXY}GXY model} 

The generalized XY (GXY) model is a spin nematic deformation of the standard XY model
\cite{1985Korshunov}. The Hamiltonian (with finite external field) is given by:
\begin{equation}
\label{eq:gXY_ham} 
\mathcal{H} = -\Delta \sum_{\langle ij \rangle} \cos(\theta_i - \theta_j) - (1-\Delta)  \sum_{\langle ij \rangle} \cos(2(\theta_i - \theta_j)) - h \sum_i \cos\theta_i, 
\end{equation}
\begin{figure}
	\centering 
	\includegraphics[width=0.8\textwidth]{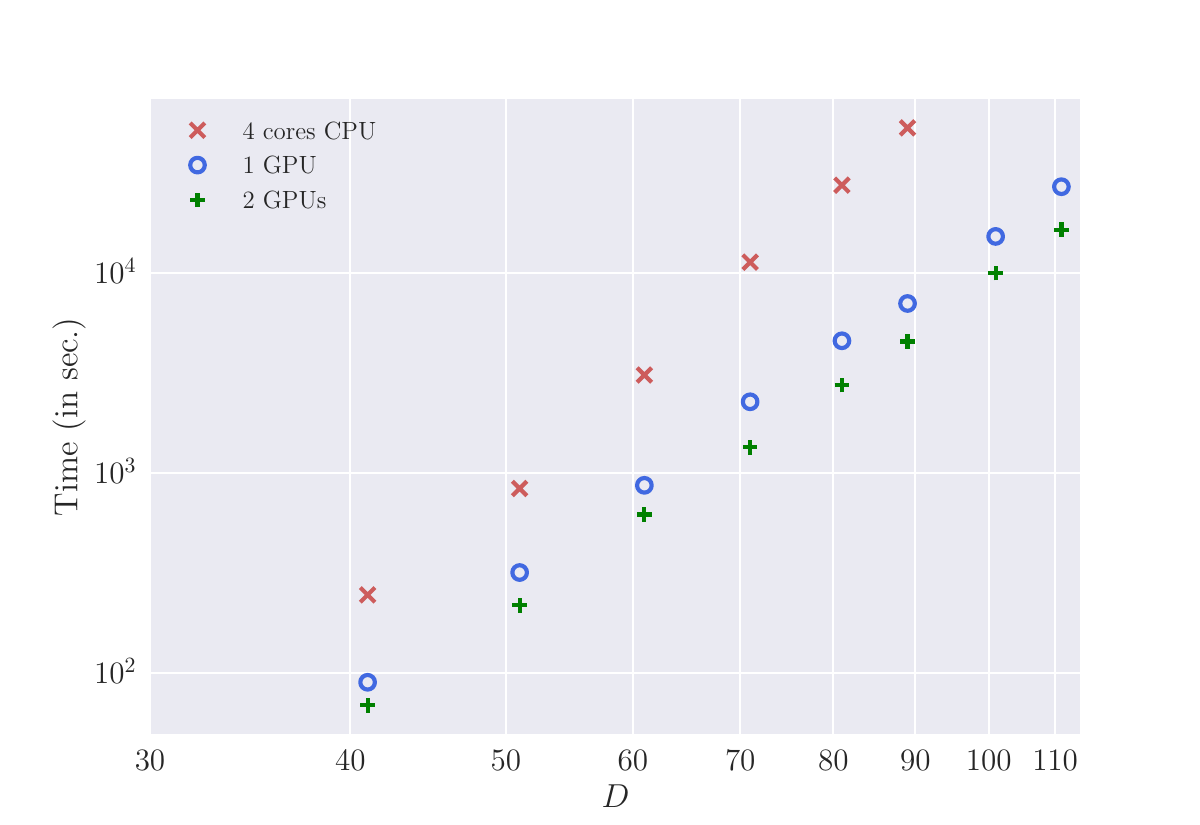}
	\caption{\label{fig:timeCPU-GPU}The runtime in seconds for the GXY model with different $D$ on lattice of size $2^{30} \times 2^{30}$ with CPU version and \TT. We used maximum $D=111$ with \TT and $D=89$ with the standard CPU version to compare the timings. \CHANGE{Note that these timings are just for the computation of $Z$ without any impure tensor computations. If we insert impure tensors (say to compute magnetization) then for $D=101$, the run time on a single GPU increases to $\sim 19,000$ seconds compared to $15,200$ seconds for a pure network.}} 
\end{figure}
where we follow the standard notation $\langle ij \rangle$ to denote nearest neighbours and $\theta_{i} \in [0,2\pi)$. Two limits are clear cut: $\Delta = 0$ corresponds to a pure spin-nematic phase and $\Delta = 1$ is the standard XY model. We will report on the ongoing tensor formulation of this model in a separate work \cite{Jha:2023abc}. 

%The reason we consider this model to compare the run time on CPU and with CUDA is that the effect of truncation (i.e., finiteness of $D$) is injected into the TRG from the beginning even before we start to coarse-grain the system because of the continuous global $O(2)$ symmetry. 

\CHANGE{To test our GPU-acceleration, we have focused on two models. One with discrete symmetry (Ising model) and the other with continuous global symmetry (GXY). Since the indices for the initial tensor in GXY model takes infinite values, a suitable truncation is needed from the first step of coarse-graining while for the Ising model, the first few iterations can be carried out exactly due to the size of the initial tensor.}

For the GXY model in this work, we will only consider $h=0$. We performed tensor computations for a fixed value of $\Delta=0.5$ and for different $D$. The computation time for this model scaled like $\sim D^{5.4(3)}$ with \TT~ while the CPU timings were close to $\sim D^{7}$ which is consistent with the expectation of higher-order TRG scaling in two dimensions. \CHANGE{Note that since the cost of SVD scales like $\sim D^{6}$, the observed computation time complexity also signals the fact that even SVD computations are accelerated by GPU, though they are sub-dominant compared to the tensor contractions.}

We show the comparison between the run times showing the CUDA acceleration with \TT in Fig.~\ref{fig:timeCPU-GPU}. We used one and two CUDA devices available with NVIDIA GeForce RTX 2080 Ti. We found that the latter is a factor of about $\texttt{1.5x}$ faster. 
In addition, we also tested our program on 4 CUDA devices with NVIDIA TITAN RTX\footnote{We thank Nobuo Sato for access to the computing facility.} and found a further speedup of $\sim \texttt{1.3x}$ (not shown in the figure) over two CUDA RTX 2080 Ti for range of $D$. 
Therefore, it is clear that with better GPU architectures in the future, TRG computations will benefit significantly from moving over completely to GPU-based computations.

\begin{figure}
	\centering 
	\includegraphics[width=0.7\textwidth]{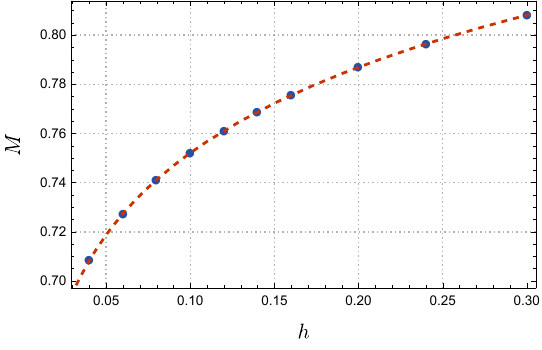}
	\caption{\label{fig:delta} The magnetization of the XY model at various finite magnetic fields $h$ at the critical temperature on lattice size of $2^{36} \times 2^{36}$. The fit is $0.877(3) h^{1/15.11(17)}$.}  
\end{figure}

\CHANGE{Using our GPU-enhanced code, we can access larger $D$ and there is possibility of computing the critical exponents accurately. We fixed $\Delta=1$ in Eq.~\ref{eq:gXY_ham} and computed the critical exponent
$\delta$ defined as $M \sim h^{1/\delta}$ at $T = T_{c} = 0.89290$ \cite{Jha:2020oik} and obtained $\delta = 15.11(17)$ compared to the exact result of 15. The results are shown in Fig.~\ref{fig:delta} with $D=111$. To ensure convergence in the bond dimension, we also took $D=131$ and found that difference in magnetization between the two $D$ is $ \sim 10^{-8}$.}

%---------------------------
\subsection{Ising model} 

\begin{figure}
	\centering 
	\includegraphics[width=1.0\textwidth]{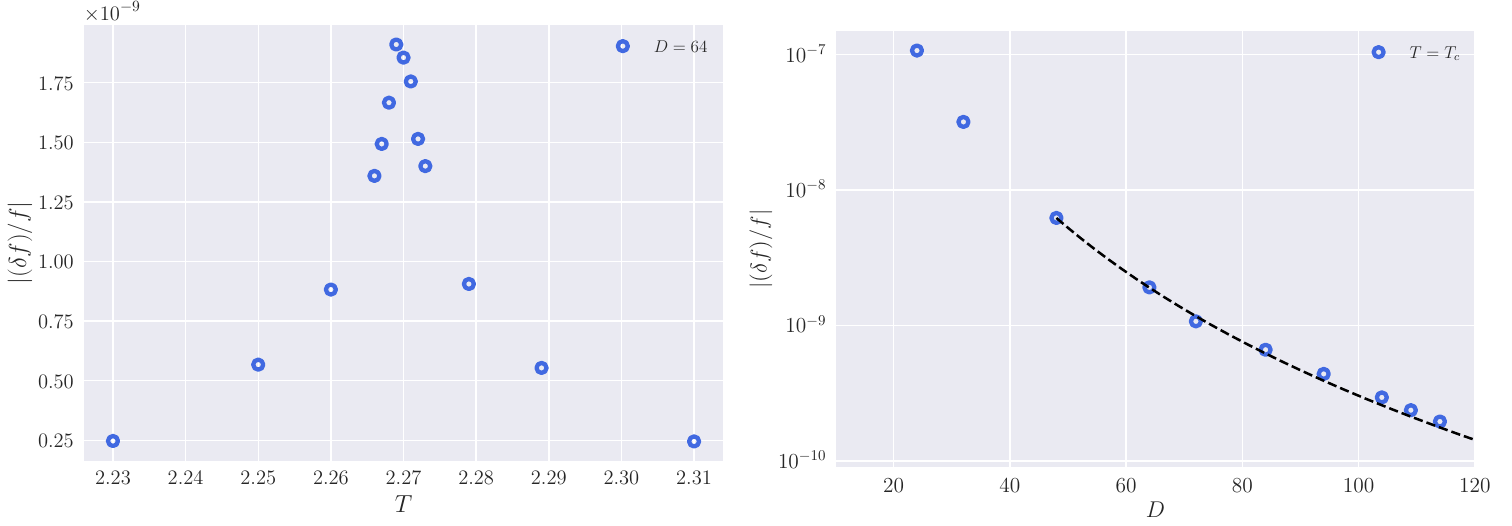}
	\caption{\label{fig:ising1}\textbf{Left}: The deviation of the TRG results from the exact result (\ref{eq:isingEX}). \textbf{Right}: The dependence of the error on $D$ and therefore on the execution time at $T = T_{c}$.}  
\end{figure}

In the previous subsection, we compared the run time on the GXY model, however, we also want to test the algorithm with \code{opt\_einsum} and GPU acceleration on a model with a known solution. In this regard, we considered the Ising model on a square lattice which admits an exact solution. This makes it a good candidate for the sanity check of the algorithm and the convergence properties. We will check the accuracy of the higher-order TRG method by computing the free energy which is the fundamental quantity accessible in TRG computations. It can be obtained directly in the thermodynamic limit from the canonical partition function $Z$ as $-T\ln Z$. The exact result for the free energy of the Ising model is given by:
\begin{equation}
\label{eq:isingEX} 
    f_{\rm{E}} = - \frac{1}{\beta} \Bigg(\ln (2 \cosh (2 \beta))- \kappa^{2} ~ 
\pFq{4}{3}{1,1,\frac{3}{2},\frac{3}{2}}{2,2,2}{16 \kappa ^{2}}\Bigg),
\end{equation}
where $\kappa = \sinh (2\beta)/(2\cosh^2{2\beta})$ and $_{p}F{_q}$ is the generalized hypergeometric function and $\beta$ is the inverse temperature. We define the error in TRG computation of the free energy as:
\begin{equation}
    \Big \vert  \frac{\delta f}{f}  \Big \vert = 
    \Big \vert \frac{f_{\text{TRG}} - f_{\text{E}}}{f_{\text{E}}} \Big \vert. 
\end{equation}
We show the results for this observable for various $T$ in the left panel of Fig.~\ref{fig:ising1} and at fixed $T=T_{c}$ for various $D$ (run time) in the right panel of Fig.~\ref{fig:ising1}. Each data point in the left panel of Fig.~\ref{fig:ising1} took about $2000$ seconds on 4 cores of Intel(R) Xeon(R) Gold 6148. The largest deviation we observed (as expected) was at $T = T_{c} \sim 2.269$ where $\vert (\delta f)/f \vert$ was $1.91 \times 10^{-9}$. We could not find any other algorithm with such accuracy for the same execution time. Note that we did not even use the CUDA acceleration for this comparison. We used a bond dimension of $D=64$ and computed the free energy on a square lattice of size $2^{20} \times 2^{20}$. In order to ensure that the result has converged properly, we also studied lattice size $2^{25} \times 2^{25}$ and obtained the same deviation from the exact result. 

\begin{table}[htbp]
    \centering
    \begin{tabular}{cccccc}
        $D$ & $\Big \vert  \frac{\delta f}{f}  \Big \vert$ &  A100  & RTX 2080 & 4 CPUS \\
        \\ \hline 
        84   & $6.6 \times 10^{-10}$ & 6004 & 9171 & 11714 \\ 
        \hline 
        94   & $4.4 \times 10^{-10}$ & 11960 & 19305 & 29376 \\
        \hline
        104 & $2.9 \times 10^{-10}$ & 21376 & 36159 & 58715 \\
        \hline
        109 & $2.4 \times 10^{-10}$ & 28942 & 46350 & 80578 \\
        \hline
    \end{tabular}
    \caption{Timings (in seconds) for the HOTRG algorithm for Ising model for $2^{20} \times 2^{20}$ lattice at $T=T_c$ using CPU and GPU.} 
\end{table}

Another useful quantity to compute is the coefficient $\alpha$ defined as $\vert (\delta f)/f \vert \propto 1/D^{\alpha}$. Different TRG algorithms have different $\alpha$ and a higher value represents faster convergence with the bond dimension $D$. We 
show the error as a function of $D$ at $T = T_{c}$ for Ising model in the right panel of Fig.~\ref{fig:ising1}. Doing a fit of the numerical fit of the data for $ D \in [48, 114]$ gave $\alpha = 4.12(2)$. \CHANGE{This exponent has close relation to the scaling of entanglement entropy of the critical $c=1/2$ Ising model which in turn is related to the finite $D$ scaling of correlation length given by
$\xi_{D} \sim D^{\kappa}$ \cite{PhysRevB.78.024410}. It is argued that the error in free energy scales like $\sim \xi^{-d}$ for $d=2$ dimensional Ising model with maximum error at $T = T_{c}$ where the correlation length diverges. This implies that $\vert (\delta f)/f \vert \propto D^{-2\kappa}$. Using this we get $\kappa \sim 2.06$ which is very close to the expected\footnote{We thank the referee for this suggestion} exponent $\kappa \sim 2.03425$}. 
For this model, we also compared our numerical results with two other recent works. The triad second renormalization group introduced in \cite{Kadoh:2021fri} can only get to an accuracy of $10^{-9}$ at $T=T_{c}$ with about $10^{5}$ seconds of CPU time which roughly translates to our CPU code being about $30$ times faster to get the same accuracy for the Ising model. The $\partial$TRG method of Ref.~\cite{Chen_2020} does not have an accuracy of $10^{-9}$ at the critical temperature even with $D=64, 128$ though admittedly it behaves much better away from critical temperatures. \CHANGE{We compared the CPU and GPU code for the Ising model and the results are summarized in Table~1. We found that using NVIDIA A100\footnote{This is currently the state-of-the-art GPU used extensively to train large language models (LLM) like GPT-4.} had much better performance than RTX 2080.}
%---------------------------
\subsection{Three-state Potts model}
\begin{figure}
	\centering 
	\includegraphics[width=0.6\textwidth]{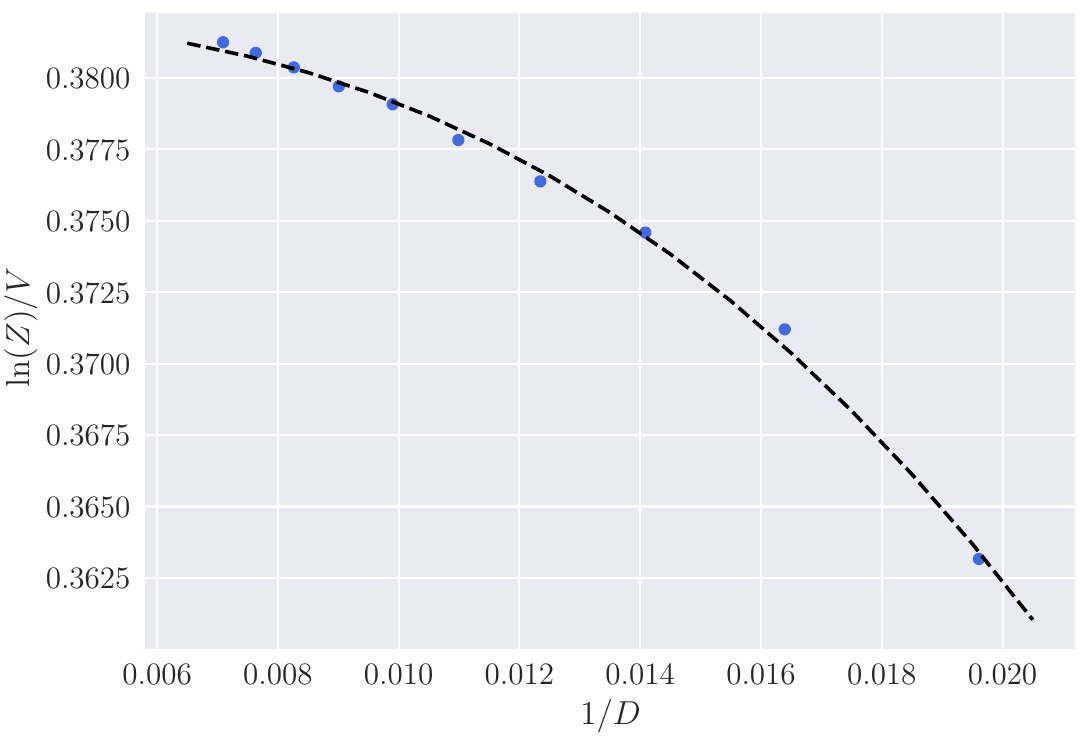}
	\caption{\label{fig:potts0}The convergence of the density of $\ln(Z)$ with inverse of bond dimensions at $T_{c}$ for the three-state Potts model on lattice of size $2^{30} \times 2^{30}$. The quadratic fit gives $\ln(Z)/V = 0.3822(2)$ in the $D\ \to \infty $ limit.} 
\end{figure}

\begin{figure}
	\centering 
	\includegraphics[width=0.7\textwidth]{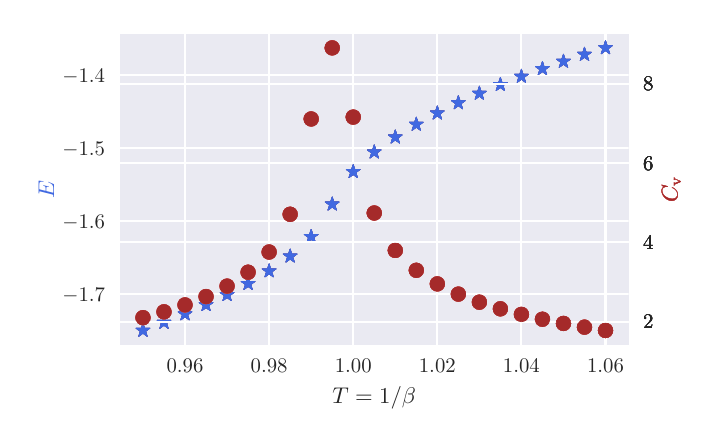}
	\caption{\label{fig:potts1}The internal energy ($E$) and specific heat ($C_{\text{v}}$) for the $q=3$ Potts model with $D=64$ on a lattice of size $2^{20} \times 2^{20}$. The continuous transition from the peak of specific heat is consistent with the exact analytical result. Each data point in the plot took about 1300 seconds using \TT on 2 CUDA devices.} 
\end{figure}

As a generalization of the Ising model, we can also consider the classical spins to take values from $1, 2, \cdots, q$. This is the $q$-state Potts model which is another widely studied statistical system. In particular, we consider the case $q=3$ as an example. On a square lattice, this model has a critical temperature that is exactly known for all $q$. The transition, however, changes order at some $q$ and the nature of the transition is continuous for $q < 4$ \cite{Wu1982}. The exact analytical result for $T_{c}$ on square lattice is:
\begin{equation}
T_{c} = \frac{1}{\ln(1 + \sqrt{q})}.  
\end{equation}
If we restrict to $q = 2$, we reproduce the Ising result (up to a factor of 2). The $q$-state Potts model has been previously considered using TRG methods both in two and three dimensions in Refs.~\cite{2010PhRvB..81q4411Z,Wang_2014,Jha:2022pgy}. The initial tensor can be written down by considering the $q \times q$ Boltzmann nearest-neighbor weight matrix as:
\begin{equation}
	\mathbb{W}_{ij} = \hspace{4mm} \begin{cases}
		e^{\beta} \hspace{3mm}  ; \hspace{4mm} \text{if $i = j$}   \quad \phantom{\infty}   \\
		1 \hspace{5mm}; \hspace{4mm} \text{otherwise}   \quad .\phantom{0}
	\end{cases}
\end{equation} 
and then splitting the $\mathbb{W}$ tensor using Cholesky factorization, i.e., $\mathbb{W} = LL^{T}$ and combining four $L$'s to make $T_{ijkl}$ as
$T_{ijkl} = L_{ia}L_{ib}L_{ic}L_{id}$. Note that this tensor can be suitably modified to admit finite magnetic fields. \CHANGE{We first study the convergence of density of 
$\ln(Z)$ with inverse bond dimension. The result is shown in Fig.~\ref{fig:potts0}}.
\CHANGE{In the absence of external magnetic field, this model has a phase transition at $T_{c} \approx 0.99497$ and we check this using \TT. The results obtained are shown in Fig.~\ref{fig:potts1}.}

\CHANGE{As mentioned before, a prime motivation of our work was to access large bond dimensions to accurately determine the critical exponents. We computed these exponents for the XY model in Sec.~\ref{subsec:GXY} and now we determine critical exponent $\alpha$ for the three-state Potts model. Taking the zero-field limit, we look at the behaviour of the specific heat as $T \to T_{c}$. Defining $\tau = (T-T_c)/T_c$, we fit the specific heat (as $\tau \to 0$)
using the standard form}: 
\begin{figure}
	\centering 
	\includegraphics[width=0.6\textwidth]{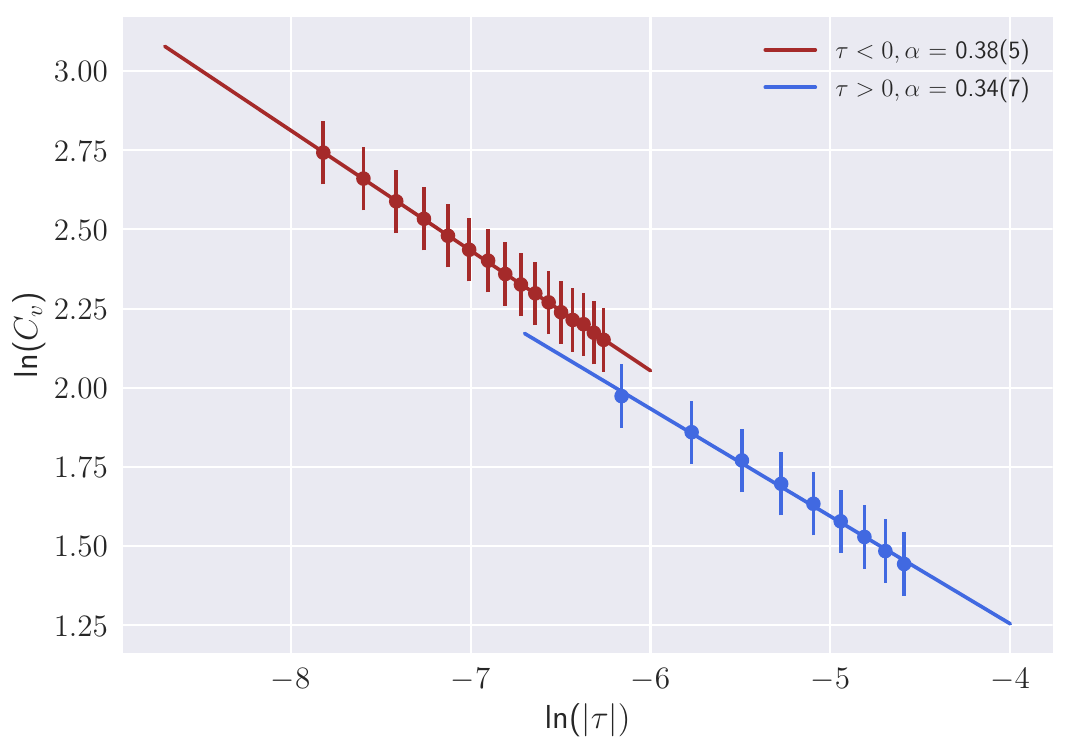}
	\caption{\label{fig:potts2}The determination of the critical exponent $\alpha$ for lattice of size $2^{30} \times 2^{30}$ with $D=111$. } 
\end{figure}
\begin{equation}
\label{eq:Cvfit} 
    \ln(C_\text{v}) = \beta - \alpha \ln \vert \tau \vert \;\; \implies \; \; C_{\text{v}} \propto \vert \tau \vert ^{-\alpha}. 
\end{equation}
\CHANGE{To compute the specific heat, we compute the second derivative of the partition function using the standard finite-difference method. We fit the numerical data with \eqref{eq:Cvfit} to obtain the critical exponent $\alpha$ taking $10\%$ error in specific heat. Using \TT, we get $\alpha = 0.38(5)$ and $0.34(7)$ respectively as shown in Fig.~\ref{fig:potts2} by fitting the data for $T < T_{c}$ and $T > T_{c}$ respectively. Our result is is consistent with the known result of $\alpha = 1/3$~\cite{Batchelor_2017}.}   
% https://iopscience.iop.org/article/10.1088/1751-8121/aa5fdc/pdf
% https://arxiv.org/abs/1307.3865
%%%%%%%%%%%%%%%%%%%%%%%%%%%%%%%%
\section{Summary}

We have described an efficient way of performing tensor renormalization group calculations with \PY using CUDA. For the two-dimensional classical statistical systems we explored in this work, there was a substantial improvement in the scaling of computation time with the bond dimension. In particular, the results show that there is \texttt{$\sim 8$x} speedup for $D=89$ for the generalized XY model on $2^{30} \times 2^{30}$ lattice using a single GPU which increases to \texttt{$\sim 12$x} using two GPUs. For a larger bond dimension of $D=105$, there is an estimated \texttt{$\sim 15$x} speedup. The scaling of computation time scales like $\sim O(D^5)$ with GPU acceleration which is to be compared with the naive CPU scaling of $\sim O(D^7)$ in two dimensions. This speedup means that one can explore larger $D$ using CUDA architecture which is often required for accurate determination of the critical exponents \CHANGE{as we have demonstrated for the XY and Potts model}. We envisage that the CUDA acceleration would also help TRG computations in higher dimensions in addition to the two (Euclidean) dimensions considered in this work. There have not been many explorations in this direction but we believe that in the future, we would see extensive use of the GPU resources. A potential bottleneck in the use of GPUs for TRG computations is the memory availability. This can often cause errors and severely limit the scope of the numerical computations. Partly due to this, we have not been able to significantly speed up any three-dimensional models yet though it appears to be possible. There are several other directions that can be pursued such as implementing a C/C++ version with \code{opt_einsum} to have better control of the available memory while utilizing the CUDA acceleration. We leave such questions for future work. % Explore Julia? :-) 

%%%%%%%%%%%%%%%%%%%%%%%%%%%%%%%%

\section*{\centering Acknowledgements}
This material is based upon work supported by the U.S. Department of Energy, Office of Science, Office of Nuclear Physics under contract DE-AC05-06OR23177. The research was also supported by the U.S. Department of Energy, Office of Science, National Quantum Information Science Research Centers, Co-design Center for Quantum Advantage under contract number DE-SC0012704. We thank the Institute for Nuclear Theory at the University of Washington for hospitality during the completion of this work. The numerical computations were done on Symmetry which is Perimeter Institute’s HPC system \CHANGE{and Syracuse University HTC Campus Grid supported by NSF award ACI-1341006.}

%%%%%%%%%%%%%%%%%%%%%%%%%%%%%%%%
\appendix  
\section{\label{sec:App}Contraction of network - Different methods}

\begin{figure}
	\centering 
	\includegraphics[width=0.65\textwidth]{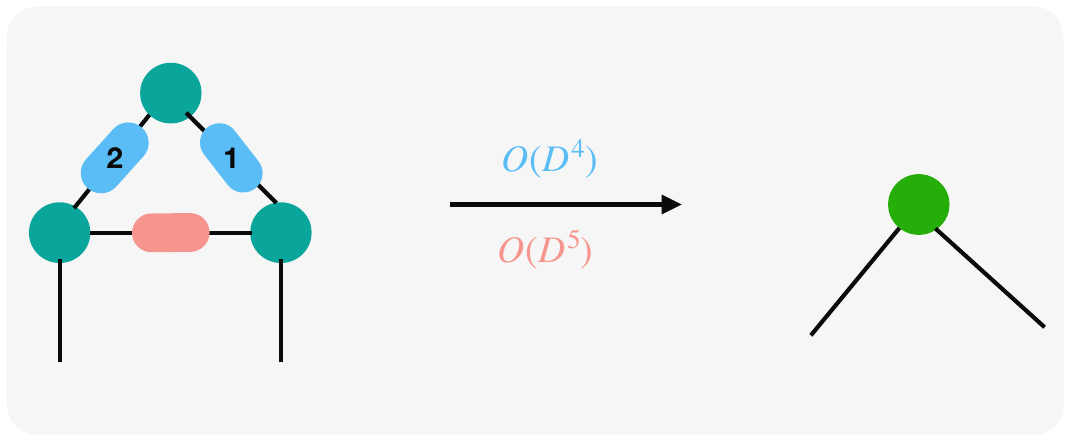}
	\caption{\label{fig:cont1}Schematic representation of two ways of contracting a network. The cost is $O(D^4)$ if we follow the order of blue-shaded regions as numbered. However, if we start by contracting the pink link first, then the leading cost will be $O(D^5)$.} 
\end{figure}

In this Appendix, we elaborate on the optimized sequence of contraction order when dealing with complicated tensor networks. In Fig.~\ref{fig:cont1}, we show two different contraction pattern that yields different time complexity. 
Let us start with two rank-three (each with $D^3$ elements) and one rank-two tensor ($D^2$ elements). Suppose we want to contract three pairs of indices and obtain a final tensor of rank-two as shown. If we follow the blue-marked regions in the order 1 and 2 as mentioned, the cost will be $O(D^4)$.  However, if we rather choose to contract the bond starting with the pink blob, then this step would be $O(D^5)$ followed by $O(D^4)$ steps leading to overall time complexity of $O(D^5)$. Hence, choosing an optimum sequence is very important for practical purposes. Fortunately, this is something \code{opt\_einsum} and \code{ncon} do fairly well. The efficient evaluation of tensor expressions that involve sum over multiple indices is a crucial aspect of research in several fields, such as quantum many-body physics, loop quantum gravity, quantum chemistry, and tensor network-based quantum computing methods \cite{Ibrahim:2022wju}. 

The computational complexity can be significantly impacted by the sequence in which the intermediate index sums are performed as shown above. Notably, finding the optimal contraction sequence for a single tensor network is widely accepted as \textbf{NP}-hard problem. In view of this, \code{opt\_einsum} relies on different heuristics to achieve near-optimal results and serves as a good approximation to the best order. This is even more important when we study tensor networks on non-regular graphs or on higher dimensional graphs. We show a small numerical demonstration below. We initialize a random matrix and set a contraction pattern option and monitor the timings. We find that all three:\code{tensordot}, \code{ncon}, \code{opt\_einsum} perform rather similarly. 
The slowest is \code{np.einsum} when the optimization flag not set (i.e., false). 
However, since we are interested in GPU acceleration in this work, we use \code{opt\_einsum} which has better support to our knowledge and is also backend independent. We also compare its performance for a specific contraction on CPU and with \code{torch} on CUDA.   

\begin{lstlisting}[language=Python,upquote=true]
import numpy as np
from opt_einsum import contract
from ncon import ncon

i, j, k, l = 80, 75, 120, 120
A = np.random.rand(i, j, k, l)
B = np.random.rand(j, i, k, l)

%timeit np.tensordot(A, B, axes=([1,0], [0,1]))
# 2.72 s \pm 38.3 ms per loop

%timeit np.einsum('ijkl,jiab->klab', A, B)
# WARNING: Never use this without optimization. 
# Slower by factor of 500x or so! Not considered.
# We can turn the optimize flag as below. 

%timeit np.einsum('ijkl,jiab->klab', A, B, optimize=True)
# 2.75 s \pm 40.2 ms per loop

%timeit contract('ijkl,jiab->klab', A, B)
# 2.69 s \pm 40.9 ms per loop

%timeit ncon((A, B),([1,2,-1,-2], [2,1,-3,-4]))
# 2.37 s \pm 20 ms per loop 

i, j, k, l = 200, 100, 80, 80
A = np.random.rand(i, j, k, l)
B = np.random.rand(j, i, k, l)

import torch
from opt_einsum_torch import EinsumPlanner
ee = EinsumPlanner(torch.device('cuda:0'), cuda_mem_limit = 0.7)

%timeit contract('ijkl,jiab->klab', A, B)
# 6.57 s \pm 80.7 ms per loop [on CPU] 

%timeit ee.einsum('ijkl,jiab->klab', A, B)
# 3.76 s \pm 16.9 ms per loop [on CUDA] 
# For this single contraction, we see a factor of about 1.7! 
\end{lstlisting}

\bibliographystyle{utphys}
%%%%%%%%%%%%%%%%%%%%%%%%%%%%%%%%%%%%%%%%
\raggedright
\bibliography{cuda_v2.bib}
\end{document}

%% file: styles.tex
\definecolor{halfgray}{gray}{0.55}
\definecolor{pyframe}{RGB}{207, 207, 207}
\definecolor{pybackground}{rgb}{0.95,0.95,0.95}
\definecolor{pypurple}{RGB}{170, 94, 255}

\definecolor{pykeyword}{HTML}{2A9D8F}
\definecolor{pyblue}{HTML}{4062BB}
\definecolor{ipython_red}{HTML}{52489C}
\definecolor{pycyan}{HTML}{E76F71}

\lstdefinelanguage{Python}{
    morekeywords={access,and,as,assert,break,class,continue,def,del,elif,else,except,exec,finally,for,from,global,if,import,in,is,lambda,not,or,pass,print,raise,return,try,while},
    morekeywords=[2]{True,False,self},
    sensitive=true,
    morecomment=[l]\#,
    morestring=[b]',
    morestring=[b]",    
    morestring=[s]{'''}{'''},
    morestring=[s]{"""}{"""},
    morestring=[s]{r'}{'},
    morestring=[s]{r"}{"},
    morestring=[s]{r'''}{'''},
    morestring=[s]{r"""}{"""},
    morestring=[s]{u'}{'},
    morestring=[s]{u"}{"},
    morestring=[s]{u'''}{'''},
    morestring=[s]{u"""}{"""},
    literate=
    {á}{{\'a}}1 {é}{{\'e}}1 {í}{{\'i}}1 {ó}{{\'o}}1 {ú}{{\'u}}1
    {Á}{{\'A}}1 {É}{{\'E}}1 {Í}{{\'I}}1 {Ó}{{\'O}}1 {Ú}{{\'U}}1
    {à}{{\`a}}1 {è}{{\`e}}1 {ì}{{\`i}}1 {ò}{{\`o}}1 {ù}{{\`u}}1
    {À}{{\`A}}1 {È}{{\'E}}1 {Ì}{{\`I}}1 {Ò}{{\`O}}1 {Ù}{{\`U}}1
    {ä}{{\"a}}1 {ë}{{\"e}}1 {ï}{{\"i}}1 {ö}{{\"o}}1 {ü}{{\"u}}1
    {Ä}{{\"A}}1 {Ë}{{\"E}}1 {Ï}{{\"I}}1 {Ö}{{\"O}}1 {Ü}{{\"U}}1
    {â}{{\^a}}1 {ê}{{\^e}}1 {î}{{\^i}}1 {ô}{{\^o}}1 {û}{{\^u}}1
    {Â}{{\^A}}1 {Ê}{{\^E}}1 {Î}{{\^I}}1 {Ô}{{\^O}}1 {Û}{{\^U}}1
    {œ}{{\oe}}1 {Œ}{{\OE}}1 {æ}{{\ae}}1 {Æ}{{\AE}}1 {ß}{{\ss}}1
    {ç}{{\c c}}1 {Ç}{{\c C}}1 {ø}{{\o}}1 {å}{{\r a}}1 {Å}{{\r A}}1
    {€}{{\EUR}}1 {£}{{\pounds}}1
    {^}{{{\color{pypurple}\^{}}}}1
    {=}{{{\color{pypurple}=}}}1
    {+}{{{\color{pypurple}+}}}1
    {*}{{{\color{pypurple}$^\ast$}}}1
    {/}{{{\color{pypurple}/}}}1
    {+=}{{{+=}}}1
    {-=}{{{-=}}}1
    {*=}{{{$^\ast$=}}}1
    {/=}{{{/=}}}1,
    literate=
    *{-}{{{\color{pypurple}-}}}1
     {?}{{{\color{pypurple}?}}}1,
    identifierstyle=\color{black}\ttfamily,
    commentstyle=\color{pycyan}\ttfamily,
    stringstyle=\color{ipython_red}\ttfamily,
    showstringspaces = false,
    keepspaces=true,
    showspaces=false,
    showstringspaces=false,
    rulecolor=\color{pyframe},
    frame=single,
    frameround={t}{t}{t}{t},
    framexleftmargin=6mm,
    numbers=left,
    numberstyle=\tiny\color{halfgray},
    backgroundcolor=\color{pybackground},
    basicstyle=\ttfamily\smaller[0.5],
    keywordstyle=\color{pykeyword}\ttfamily,
    keywords=[3]{EinsumPlanner, contract, einsum, ee, svd},
    keywordstyle={[3]\color{pyblue}},
}